\documentclass[10pt]{article}
\usepackage{amsmath,amssymb,amsfonts}
\usepackage{graphicx}
\usepackage{theorem}

\theorembodyfont{\upshape}
\newtheorem{theorem}{Theorem}
\newtheorem{lemma}{Lemma}[section]

\def\0{\varnothing}

\newcommand{\feta}{{\boldsymbol{\eta}}}
\newcommand{\fnull}{{\boldsymbol{0}}}

\title{Rigorous results on spontaneous symmetry breaking in a
one-dimensional driven particle system}
\author{Stefan Gro{\ss}kinsky\footnote{Statistical Laboratory,
    University of Cambridge, Cambridge CB3 0WB, UK\newline
email: stefan@statslab.cam.ac.uk, phone: +44 1223 337962}
, Gunter M. Sch\"utz\footnote{Institut f\"{u}r
    Festk\"{o}rperforschung, Forschungszentrum J\"{u}lich, 52425
    J\"{u}lich, Germany}, Richard D. Willmann$^\dagger$}

\begin{document}

\maketitle

\begin{abstract}
We study spontaneous symmetry breaking in a one-dimensional driven
two-species stochastic cellular automaton with parallel sublattice
update and open boundaries. The dynamics are symmetric with respect to
interchange of particles. Starting from an empty initial lattice, the
system enters a symmetry broken state after some time $T_1$ through an
amplification loop of initial fluctuations. It remains in the symmetry
broken state for a time $T_2$ through a traffic jam effect. Applying a
simple martingale argument, we obtain rigorous asymptotic estimates
for the expected times $<T_1>\propto L\ln{L}$ and $\ln{<T_2>} \propto
L$, where $L$ is the system size. The actual value of $T_1$ depends
strongly on the initial fluctuation in the amplification
loop. Numerical simulations suggest that $T_2$ is exponentially
distributed with a mean that grows exponentially in system size. For
the phase transition line we argue and confirm by simulations that the
flipping time between sign changes of the difference of particle
numbers approaches an algebraic distribution as the system size tends
to infinity.
\end{abstract}


\section{Introduction}

Spontaneous symmetry breaking (SSB) is associated with phase
transitions and is therefore not expected at positive temperature in
one-dimensional equilibrium systems with short-range interactions. The
underlying physical picture behind the absence of 1-d phase
transitions, viz. the unsuppressed creation of islands of the minority
phase inside a region of the majority phase (e.g. in an Ising system)
due to thermal noise, is very robust. So it came as a bit of a surprise
that in a quite simple stochastic (and hence noisy) lattice model of a
driven diffusive system with short-range interactions SSB was observed
\cite{Evans95a,Evans95}. In this so-called bridge model, two species of
particles $A,B$ move in opposite directions with rate 1 and are
injected with rate $\alpha$ and ejected with rate $\beta$ at the
boundary sites. Although the dynamical rules are symmetric with respect
to the two species, two phases with non-symmetrical steady states were found 
in a mean-field approximation. In the symmetry broken states there is a 
macroscopic excess amount of one particle species, i.e., the order parameter 
$\Delta=\rho_A -\rho_B$ measuring the difference of the average
particle densities of the
two species $A,B$ attains a non-zero value in the thermodynamic limit 
$L\to\infty$. These analytical results were confirmed by Monte Carlo 
simulations of finite systems of large size $L$.

In the limit of vanishing ejection rate ($\beta\to 0$) the existence of SSB
in this model could be established rigorously \cite{Godreche95}. It emerged
that (at least in this limit) the phases of spontaneously broken
symmetry are dynamically sustained by a traffic jam effect: The particles
of one species pile up at one end of the chain (because of the small
rate of ejection) and thus prevent the entrance of particles of the
other species, until by an exponentially rare fluctuation (i.e. with a
probability exponentially small in system size) no particles of that
species enter for sufficiently long time. Then the traffic jam
dissolves, allowing for particles of the other species to take over.
Later some other stochastic 1-d lattice gas models exhibiting SSB
were discovered \cite{Rako03,Levine04}, but the nature of the phase transition
in the bridge model has remained obscure \cite{Arndt98,Clincy01}. There is,
in fact, recent numerical evidence suggesting that one of the two
symmetry broken phases vanishes in the thermodynamic limit
\cite{Eric05,Pron06}.

It would seem natural to attack the problem of SSB from a macroscopic
viewpoint by deriving a hydrodynamic description of the lattice gas model
under Eulerian scaling. This approach indeed works for vanishing boundary 
rates \cite{Popk04a}, but fails for the general case due to the lack of a 
sufficiently general hydrodynamic theory for two-component systems in the 
presence of boundaries \cite{Popk04b}. Only partial results for some specific 
infinite multi-component systems are known 
\cite{Popk03a,Toth03,Clincy03,Popk04c,Fritz04,Toth05}.

These and other puzzles make driven diffusive two-component systems a matter
of considerable current interest, see \cite{Schutz03} for a review. In
\cite{Will05} we studied a variation of the bridge model with parallel
sublattice update. The deterministic bulk update scheme simplifies the
treatment of particle transport, while stochastic creation/annihilation
events occur at the boundaries in a similar fashion as in the original
bridge model. Thus -- while maintaining noisy dynamics -- the complexity of the
problem is reduced. This allowed us to determine the exact phase diagram
(for all values of the boundary rates) and to elucidate the mechanism that
leads from a symmetric particle configuration into a state with broken
symmetry. In this work we make rigorous the main results reported in
\cite{Will05}, which are complemented by new heuristic results on the
transition line. We also point out that earlier work on a similar
single-component system \cite{Schu93} yields a rigorous asymptotic
estimate for the residence time in a symmetry broken quasi-stationary
state of the two-component system and we present new results on the
phase transition between the symmetric and the symmetry broken phase.

In Sec.~2 we define the model and state our main results for the
symmetry broken phase. The proofs are given in Sec.~3. In Sec.~4 we
present simulation data which provide further insight in the relevant
time scales of the model. Analytical and simulation results for the
phase transition line are given in Sec.~5. We conclude in Sec.~6 with
some brief remarks.

\section{Model and results}

\subsection{Bridge model with sublattice parallel update}

The model considered here is defined on a one-dimensional lattice of
length $L$, where $L$ is an even number. Sites are either empty or
occupied by a single particle of either species $A$ or $B$, i.e.,
the particles are subject to an exclusion interaction and the
occupation numbers $\eta_A (i)$ and $\eta_B (i)$ of each site $i$ obey
$\eta_A (i)+\eta_B (i)\leq 1$. The dynamics is defined as a parallel
sublattice update scheme in two half steps.
In the first half-step the following processes take place: At site $1$
it is attempted to create a particle of species $A$ with
probability $\alpha\in [0,1]$ if the site is empty, or to annihilate a particle
of species $B$ with probability $\beta\in [0,1]$, provided the site is
occupied by such a particle:
\begin{equation}
0 \overset{\alpha}{\to} A\ , \quad B \overset{\beta}{\to} 0\ .
\end{equation}
Accordingly, at site $L$ a particle of species $B$ is created with probability
$\alpha$ and a particle of species $A$ is annihilated with probability
$\beta$:
\begin{equation}
0 \overset{\alpha}{\to} B\ ,\quad A \overset{\beta}{\to} 0\ .
\end{equation}
In the bulk, the following hopping processes occur deterministically
between sites $2i$ and $2i+1$ with $0<i<L/2$:
\begin{equation}
A0 {\to} 0A \ ,\quad 0B {\to} B0\ , \quad AB {\to} BA\ .
\end{equation}
In the second half-step, these deterministic hopping processes take
place between sites $2i-1$ and $2i$ with $0 < i \leq L/2$. Note that
the dynamics is symmetric with respect to interchange of the two particles 
species combined with space reflection.
The original bridge model \cite{Evans95a,Evans95} arises as the continuous-time
limit of this model with stochastic hopping and has the same symmetry.

\subsection{Results}

\begin{figure}
\begin{center}
  \includegraphics[width=0.4\textwidth]{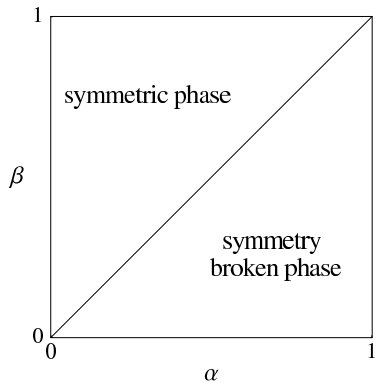}\\[2mm]
  \caption{Stationary phase diagram of the sublattice bridge
    model as a function of creation and annihilation probabilities
    $\alpha$ and $\beta$.}
\label{fig:phasediagram}
\end{center}
\end{figure}

The stationary phase diagram of the model in terms of the parameters
$\alpha$ and $\beta$ is presented in Figure~\ref{fig:phasediagram}.
The boundary lines are not difficult to analyze.  Along the line
$\alpha = 0, 0 < \beta \leq 1$ there is no injection and the stationary
state is the empty lattice (which is trivially symmetric in $A$ and $B$).
For $\beta=1, 0 \leq \alpha < 1$ it is easy to verify by direct computation
that the product measure with alternating densities
$\rho_A (i)=0$, $\rho_B (i)=\alpha/(1+\alpha)$ if $i$ is odd, and
$\rho_A (i)=\alpha /(1+\alpha)$, $\rho_B (i)=0$ if $i$ is even
is invariant. Also this stationary state is symmetric.
For $\beta=0, 0 \leq \alpha \leq 1$ there is no ejection and the
system is highly non-ergodic. Any blocking measure with $A$-particles
accumulating at the right boundary and $B$-particles
accumulating at the left boundary is invariant. We notice that most
of these measures are not symmetric. Finally, for $\alpha=1, 0 < \beta < 1$
there are two stationary product measures, one with $\rho_B (i)=0$ for
all $i$ and $\rho_A (i)=1$ for even $i$, $\rho_A (i)=1-\beta$ for odd
$i$ and an analogous one with $A$ and $B$ particles interchanged. In
each case one species
is completely expelled from the system even on a finite lattice.
This is a trivial, absorbing form of SSB with no transition between
the two states. The dynamics reduce to that of the single-species
lattice gas studied by one of us earlier \cite{Schu93}. In the special
deterministic point  $\alpha=\beta=1$
there are three invariant measures arising as limits of the measures
described above. One is the symmetric alternating $A/B$ measure
($\beta=1, \alpha\to 1$),
the other two are the symmetry-broken measures with only $A$ or
only $B$ particles ($\alpha= 1,\beta \to1$). These considerations hold
for all system sizes $L$, and in what follows the
boundary lines are excluded from the discussion.

\begin{figure}
\begin{center}
  \includegraphics[width=0.48\textwidth]{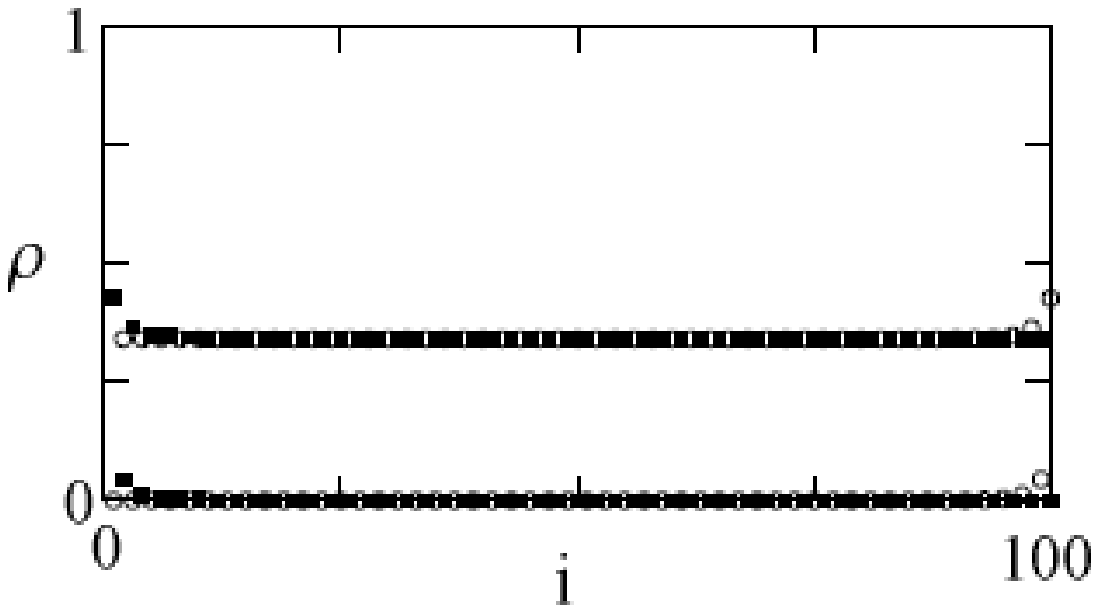}\hfill
  \includegraphics[width=0.48\textwidth]{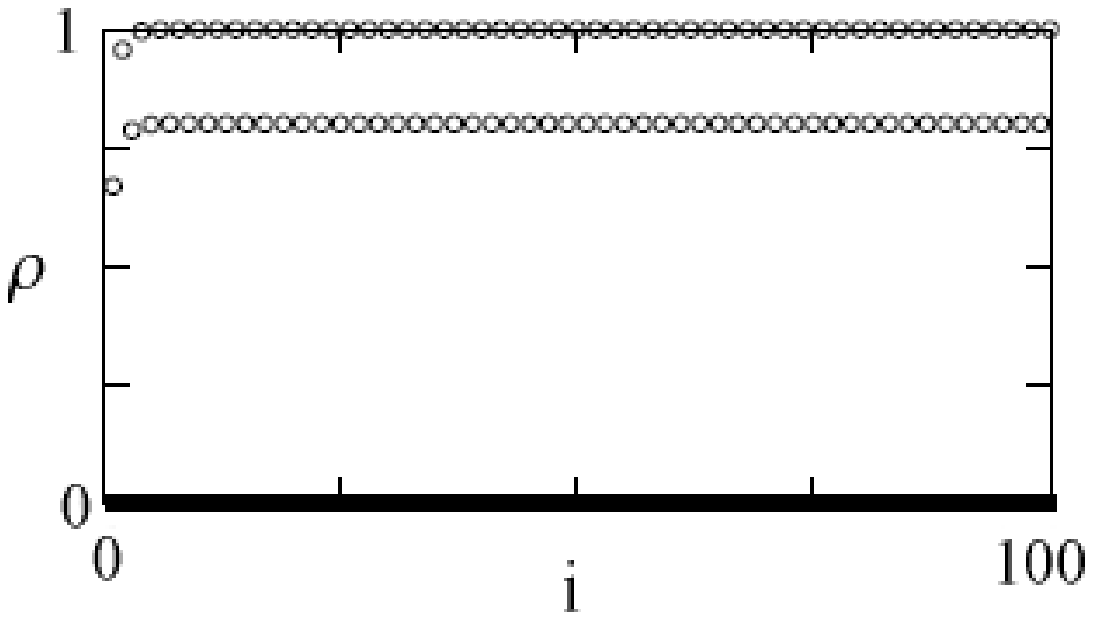}\\
  \includegraphics[width=0.48\textwidth]{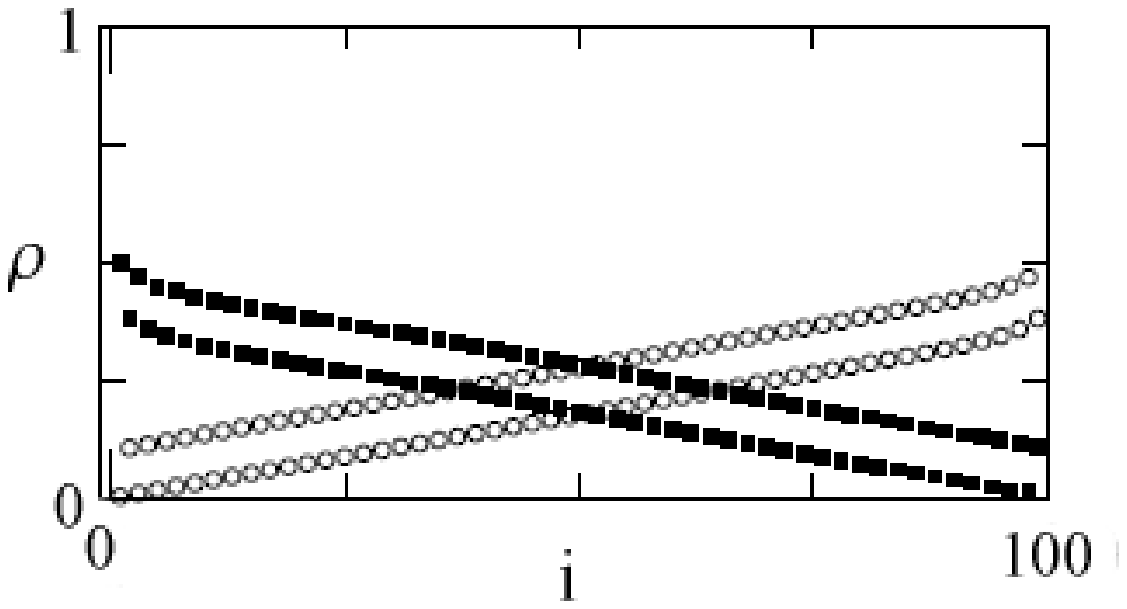}\\
\raisebox{58mm}[0mm]{\quad $\alpha =0.6$, $\beta
  =0.8$}\qquad\raisebox{25mm}[0mm]{$\alpha =\beta
  =0.2$}\qquad\quad\raisebox{50mm}[0mm]{$\alpha =0.6$, $\beta
  =0.2$}\\[-7mm]
  \caption{Average density profiles obtained from Monte Carlo
    simulations in the symmetric phase 
    (top left), the broken phase 
    (top right) and on the transition line 
    (bottom). $A$ densities are shown by $\circ$, $B$ densities
    by $\scriptstyle\blacksquare$.}
\label{fig:profiles}
\end{center}
\end{figure}

The interior of the phase diagram can be explored by Monte Carlo
simulations \cite{Will05}. Two phases are found:
\begin{itemize}
\item If $\alpha<\beta$, the system exhibits a symmetric steady
state. Here, the bulk densities in the limit $L\to\infty$ are
$\rho_A(i)=0$, $\rho_B(i)=\alpha \beta /(\alpha+\beta)$ if $i$ is odd, and
$\rho_A(i)=\alpha \beta /(\alpha+\beta)$, $\rho_B(i)=0$ if $i$ is even.
\item If $\alpha>\beta$, the system resides in the symmetry broken
  phase. Assume the $A$ particles to be in the majority. Then, the
  bulk densities in the limit $L\to\infty$ are $\rho_B(i)=0$ for all $i$,
  $\rho_A(i)=1$ for $i$ even and $\rho_A(i)=1-\beta$ for $i$ odd. This
  means that the symmetry is maximally broken and the minority
  species is completely expelled from the system.
\item On the transition line for $\alpha =\beta$ the system switches between
symmetric and broken states, which is studied numerically in Section
5.
\end{itemize}
The intention of this paper is to further elucidate the dynamics
leading to symmetry breaking which were identified in \cite{Will05}
and to determine rigorously the time scales associated with SSB for
$\alpha>\beta$.

If one of the species is expelled from the system, the dynamics
of the majority species for $\alpha >\beta$ is identical
to the single species ASEP with parallel sublattice update. The exact
steady state distribution $\mu$ for this system has been characterized in
\cite{Schu93} for all $\alpha$, $\beta$ and all system sizes $L$. Let
$\delta^1_\fnull$ be the distribution that concentrates on the empty
lattice for the single species system. Then we write
\begin{equation}
\nu_A =\mu\otimes\delta^1_\fnull\quad\mbox{and}\quad\nu_B
=\delta^1_\fnull \otimes\mu
\end{equation}
for the distributions of our two species system, where one of the
species is expelled and the other distributed according to $\mu$. Note
that we do not explicitly write the dependence of $\nu_A$ and $\nu_B$
on $\alpha$, $\beta$ and $L$. We say that the system reaches a
configuration $\feta =\big(\eta_A (i),\eta_B (i)\big)_{i=1,..,L}$ with
broken symmetry if $\eta_A (i)=1$ for all even $i$ and $\eta_B (i)=0$
for all $i$, or vice versa $\eta_B (i)=1$ for all odd $i$ and $\eta_A
(i)=0$ for all $i$. Let $T_1 (\feta )$ be the (random) time when the
system first reaches a symmetry broken configuration, starting from
configuration $\feta$. $T_2 (\feta )$ is defined to be the time until
the first particle of species $B$ enters the system.
With these definitions we can state the main results of this paper.

\begin{theorem}\label{theo1}
\ \ \textbf{Time to reach a symmetry broken configuration}\\[2mm]
For $\alpha >\beta$ we have\quad $\displaystyle\limsup_{L\to\infty}
\frac{\langle T_1 \rangle_{\delta_\fnull}}{L\ln L} <\infty$\ .
\end{theorem}

Here $\langle ..\rangle_\nu$ denotes the expected value over the time
evolution of the process with initial distribution $\nu$. So if we
start the process with an empty lattice ($\delta_\fnull
=\delta^1_\fnull \otimes\delta^1_\fnull$) then the expected time to
reach a symmetry broken state is not growing faster than $L\ln L$ with
the system size $L$.

\begin{theorem}\label{theo2}
\ \ \textbf{Stability of states with broken symmetry}\\[2mm]
For $\alpha >\beta$ we have\quad
$\displaystyle\liminf_{L\to\infty}\frac{\ln \langle T_2
  \rangle_{\nu_A}}{L}>0$\ .
\end{theorem}

So starting with a symmetry broken state $\nu_A$ where the $B$
particles are expelled from the system, the expected time until the
next $B$ particle enters grows exponentially with $L$. By symmetry an
analogous statement holds if $A$ and $B$ particles are interchanged.

Since $\nu_A =\mu\otimes\delta^1_\fnull$ and $\mu$ is stationary for
the single species system, $\nu_A$ is invariant for the two species
system for times $t<T_2$. So although $\nu_A$ and $\nu_B$ are not
stationary for finite $L$, they are exponentially stable by Theorem 2
for $\alpha >\beta$. Together with Theorem 1 this provides a proof for
spontaneous symmetry breaking in this model in the interior region of
the phase diagram. Note that this argument is done without knowing the
stationary distribution exactly, which should of course be close to
the mixture $\frac12\nu_A +\frac12\nu_B$, corresponding to the usual
concept of a phase transition in the context of Gibbs measures (e.g.\
in an Ising system).

The proofs are given in Section 3, where the proof of Theorem 2 is a
straightforward application of the results in \cite{Schu93}. The proof
of Theorem 1 relies on a simple martingale argument for an interesting
amplification mechanism which has been published in \cite{Will05}, and
will be explained again in the next subsection for self-containedness.

\subsection{Dynamics of symmetry breaking}

It is assumed that at $t=0$ there are no particles in the system and
that $\alpha>\beta >0$. Other symmetric initial conditions with a non-empty
lattice can be treated in a similar fashion.
 Starting from the empty lattice, $A$ $(B)$ particles are created
at every time step with probability $\alpha$ at site $1$ $(L)$. Once
injected, particles move deterministically with velocity $2$
$(-2)$. Therefore, at time $t=L/2$ the system is in a state where
the density of $A$ $(B)$ particles is $\alpha$ $(0)$ at all even sites and $0$
$(\alpha)$ at all odd sites.
In this situation both creation and annihilation of particles are
possible.

However, it turns out that the effect of creation of particles
is negligible:
Since $\alpha >\beta$ the deterministic hopping with velocity $2$
transports on average more $A$-particles
towards site $L$ than can be annihilated there. This leads to the formation of
an $A$-particle \textbf{jam} at the right boundary, blocking the injection of
$B$-particles. An analogous argument holds for the left boundary,
which is blocked by a $B$-particle jam. In these jams, the only source
of vacancies is annihilation at the boundaries with probability
$\beta$ in the first half-step. In the second half-step the vacancy
moves one site towards the bulk with probability $1$. Therefore, in a jam, the
density of $A$ $(B)$ particles at even (odd) sites is 1, while that at
odd (even) sites is $1-\beta$. So the only way to create particles in
this situation is a complete dissolution of a jam. But as long as it
gains particles from transport through the bulk this is a very rare event
since $\alpha >\beta$. We show in Lemma \ref{lemma2} (Section 3) that
the average number of created particles is small and bounded
independent of $L$. So creation of particles in this jammed situation
becomes negligible in the limit $L\to\infty$ and will be neglected in
the following explanation.

\begin{figure}
\begin{center}
\hspace*{5mm}\includegraphics[width=0.4\textwidth]{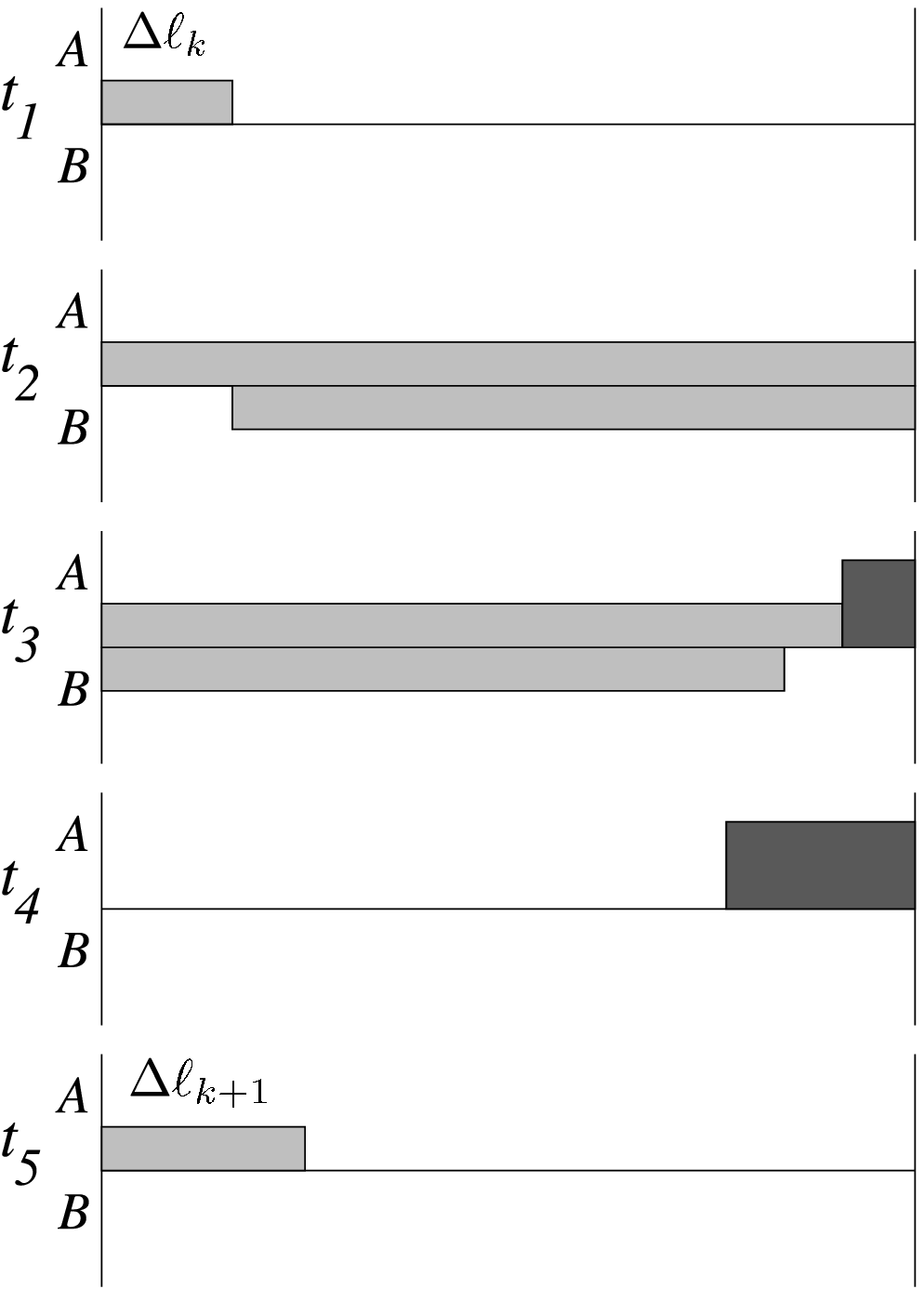}\hfill\raisebox{10mm}
{\includegraphics[width=0.5\textwidth]{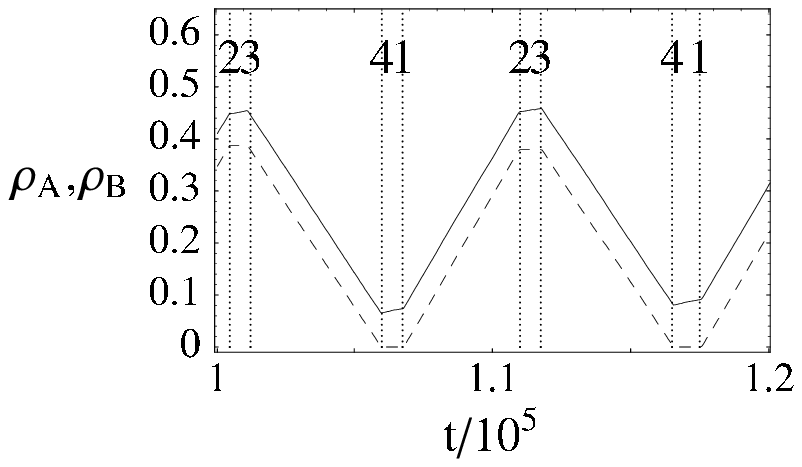}}\\[15mm]
\includegraphics[width=0.45\textwidth]{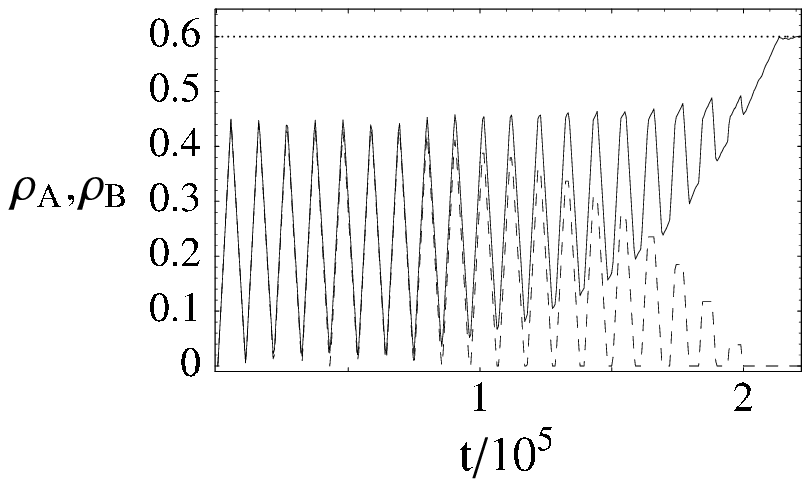}\hfill\includegraphics[width=0.45\textwidth]{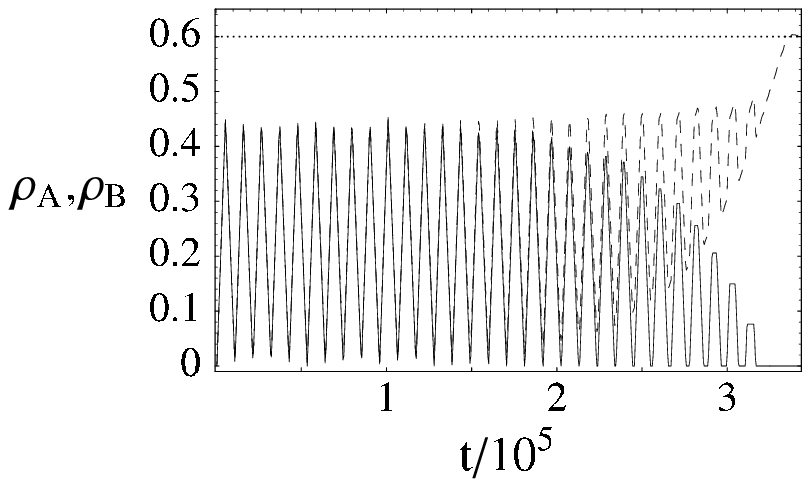}
\caption{
Cyclic behavior of the dynamics of symmetry breaking.\newline
Top: Illustration of the stages involved in the $k$-th cycle of the 
amplification loop as explained in the
    text. Low density regions are drawn light grey, jams are
    dark grey and white regions of the system are empty. On the right the 
stages are identified in a blow-up of simulation data shown in the
bottom.\newline
Bottom: Two realizations of MC simulation of symmetry breaking, starting from 
the empty lattice with $\alpha=0.9$, $\beta=0.8$ and $L=10000$. The density 
$\rho_A$is drawn in a full line (---), $\rho_B$ in dashed 
line (- - -) and the dotted line indicates the density in the symmetry
broken state.
}
\label{fig:stages}
\end{center}
\end{figure}

The number of particles in each of the two jams reduces by one in
every time step with probability $\beta$. Since creation of
particles is negligible, the influx into the jam
ceases after some time and the jam eventually dissolves. By
fluctuations, one of the jams, say the $B$-jam at the left boundary,
dissolves first. $A$ particles can enter the system while $B$
particles are still blocked until the $A$-jam at the right boundary is
also dissolved.
The configuration of the system at this point is illustrated in Figure
\ref{fig:stages} at time $t_1$ (setting $k=1$). The light grey region denotes 
a
\textbf{region of low density} of $A$ particles where the density is
$\alpha$ $(0)$ on even (odd) sites. The (random) length of this
region, $\Delta\ell_1$, describes the majority of one of the
species. The description is symmetric, so if $\Delta\ell_1 <0$ this
corresponds to a majority of $B$ particles.
Thus the average value $\langle\Delta\ell_1 \rangle_{\delta_\fnull}
=0$, but typically $\Delta\ell_1 =O(\sqrt{L})$ due to fluctuations for
large $L$ and one of the species has the majority (see Lemma
\ref{lemma3} in Section 3).

The time evolution just described
constitutes the first cycle of a periodic behavior which can be effectively
described by the dynamics of low density regions and jams at the
boundaries. The key ingredient for this simplification is the jamming
mechanism described above. The cyclic behavior consists of 4 stages,
which we summarize in the following and which are illustrated
in Figure \ref{fig:stages}. For simplicity of presentation we assume that 
$\Delta\ell_k \geq 0$, i.e.\ the $A$ particles have the majority.

\begin{enumerate}
\item At the beginning of a cycle ($t_1 :=0$) there is a low density
  region of $A$ particles at the left boundary of length
  $|\Delta\ell_k |\geq 0$.

Both species enter the system with probability
  $\alpha$ and penetrate the bulk deterministically with speed $2$ $(-2)$.

\item The low density region of $A$ particles reaches the right boundary at
  time $t_2 :=\big( L-|\Delta\ell_k |\big) /2$, blocking the creation of $B$ 
particles.

$A$ particles still enter with
  probability $\alpha$ and exit with probability $\beta <\alpha$,
  further increasing their majority.

\item At time $t_3 :=L/2$ the $B$ particles reach the left boundary,
  blocking also the creation of $A$ particles.

Both species form jams
  at the boundaries, which gain particles from the low density regions.
  Since creation of particles at the boundaries is negligible, both
  jams eventually dissolve.

\item Let $t_4$ be the time when the jam of $B$ particles is dissolved.

$A$ particles start to enter the system. Again, since $\alpha
  >\beta$ the majority of $A$ particles increases on average.

\item At time $t_5$ the $A$-jam at the right boundary is dissolved and also 
$B$ particles can enter the system.
\end{enumerate}
The cycle is finished when both jams are dissolved, i.e.\ at time 
$\max\{ t_4 ,t_5 \}$. Note that for given $\Delta\ell_k$, $t_1$, $t_2$
and $t_3$ are deterministic, whereas $t_4$ and $t_5$ are random times,
which will be defined more precisely in Section 3. In Figure
\ref{fig:stages} it is assumed that the $B$-jam dissolves first,
i.e.\ $t_4 <t_5$, which is most likely if $\Delta\ell_k >0$. But in
general $t_4 \geq t_5$ is also possible and included in the above
description. The result of the cycle is
\begin{equation}\label{result}
\Delta\ell_{k+1} :=2\, (t_5 -t_4 )\ ,
\end{equation}
which is the initial condition for the next cycle. As long as
$-L<\Delta\ell_{k+1} <L$ the cycle can restart with $0<t_2 <t_3$, and
all the stages are well defined. Note that within this framework the
process starting from the empty lattice is a cycle with
initial condition $\Delta\ell_0 =0$ and $t_2 =t_3$.
The cyclic behavior leads to an amplification of the initial
fluctuation $\Delta\ell_1$, namely
\begin{equation}
\langle\Delta\ell_k |\Delta\ell_1 \rangle_{\delta_\fnull}\simeq
\big(2\alpha /\beta -1\big)^k \Delta\ell_1\qquad\mbox{for large }L\ ,
\end{equation}
as given in Lemma \ref{lemma2}. Together with the fact that the
average duration of a cycle is of order $L$ (see Lemma \ref{lemma3}),
this is the core of the proof of Theorem \ref{theo1} provided in the
next subsection.

\section{Proofs}
  \subsection{Proof of Theorem 1}

In order to define the random times $t_4$ and $t_5$ in the cycle described 
above more precisely, we use the following procedure: At time $t_2$ we mark
the last particle of the minority species, corresponding to the rightmost $B$ 
particle in the above description. If due to a fluctuation there is no 
particle of that species in the system, the cycle is finished at $t_2$ with 
$\Delta\ell_{k+1} =L$ and also the amplification loop stops. Otherwise, we 
analogously mark the last particle of the majority species at time $t_3$ and 
define $t_4$ and $t_5$ to be the time when the marked $B$ and $A$ particle, 
respectively, just left the system. If there were no particles of the 
majority species in the system at time $t_3$, we set $t_5 := t_3$ (or $t_4 := 
t_3$ if the $B$ particles have the majority). At time $\max\{ t_4 ,t_5 \}$ we 
restart the cycle with initial condition $\Delta\ell_{k+1}$ given in 
(\ref{result}). This determines a process $(\Delta\ell_k )_{k=0,1,..}$ on 
$\mathbb{Z}$ with $\Delta\ell_0 =0$. If $|\Delta\ell_{k+1} |\geq L$ the 
process stops and we reach a symmetry broken state.

We first note some Lemmas used in the proof. The proofs of the Lemmas are
given in the next subsection.

\begin{lemma}\label{lemma1}
If $|\Delta\ell_k |\geq L$ the system reaches a symmetry broken
configuration as defined in Section 2.2 within a time of order $L$.
\end{lemma}

We define $\bar k$ to be the (random) number of cycles performed when the loop 
stops, i.e.
\begin{equation}
\bar k:=\inf\big\{ k\, :\, |\Delta\ell_k |\geq L\big\}\ .
\end{equation}
This is a stopping time for the process $(\Delta\ell_k )_{k=0,1,..}$ and in 
the following we aim to find an estimate on the expected value $\langle\bar 
k\rangle$. We will use the following recursion relation:

\begin{lemma}\label{lemma2}
For a single cycle
\begin{equation}\label{cycle}
\big\langle \Delta\ell_{k+1} \,\big|\,\Delta\ell_k
\big\rangle = q\,\Delta\ell_k +C_k \ ,
\end{equation}
where $q=\big( 2\,\frac\alpha\beta -1\big) >1$ and the constants $C_k$ are 
bounded independent of $L$ and $k$. This constitutes an amplification of the 
initial fluctuations $\Delta\ell_1$ and we get inductively
\begin{equation}\label{amp}
\big\langle\Delta\ell_{k+1} \,\big|\,\Delta\ell_1 \big\rangle=
\Delta\ell_1 q^k +\sum_{i=1}^k C_i q^{k-i}\ .
\end{equation}
\end{lemma}

Here and in the following we omit the subscript $\delta_\fnull$ since
all expectations are understood with respect to the empty initial
condition.

\begin{lemma}\label{lemma3}
The average length of a cycle is bounded above by
\begin{equation}\label{looplength}
\langle t_5 |\Delta\ell_1 \rangle =\Big(\frac\alpha\beta +1\Big)\frac L2
+\Big(\frac\alpha\beta -1\Big)\, \big\langle\Delta\ell_k
\,\big|\,\Delta\ell_1 \big\rangle /2 =O(L)\ .
\end{equation}
and the initial fluctuation is of order $\sqrt{L}$ since
\begin{equation}\label{l1}
Var (\Delta\ell_1 )=4L\frac{\alpha (2-\alpha -\beta )}{\beta^2}\ .
\end{equation}
\end{lemma}

To get an estimate for $\langle\bar k\rangle$ we use the optional stopping 
theorem for martingales. Since $q>1$ and $\Delta\ell_1 =O(\sqrt{L})$ as shown 
in (\ref{l1}), it is clear from (\ref{cycle}) that $(\Delta\ell_k )_k$ 
is a sub-martingale for sufficiently large $L$. Conditioned on the initial 
value $\Delta\ell_1$ we define the process 
$(Y_k )_{k=1,2,..}$ by $Y_1 :=\Delta\ell_1$ and
\begin{equation}
Y_k -Y_{k-1}:=\Delta\ell_k -\big\langle \Delta\ell_k \,\big|\,\Delta\ell_{k-1} 
\big\rangle\quad\mbox{for }k\geq 2\ ,
\end{equation}
following Doob's decomposition of sub-martingales. By construction, $(Y_k )_k$ 
is a martingale and $\bar k$ is a stopping time for $(Y_k )_k$. Thus by 
optional stopping and using Lemma \ref{lemma2} we have for $k\geq 2$
\begin{eqnarray}
0&=&\langle Y_{\bar k} |\Delta\ell_1 \rangle -\langle Y_{\bar k-1}
|\Delta\ell_1 \rangle =\nonumber\\
&=&\langle \Delta\ell_{\bar k} |\Delta\ell_1 \rangle -q\langle 
\Delta\ell_{\bar k-1} |\Delta\ell_1 \rangle -\langle C_{\bar k-1}|
\Delta\ell_1\rangle \ .
\end{eqnarray}
Also by Lemma \ref{lemma2} we know that $\big|\langle \Delta\ell_{\bar
  k} |\Delta\ell_1 \rangle\big|\leq qL+O(1)$, since otherwise, the
  process would have stopped before $\bar k$. Using (\ref{amp}) and
  $\langle C_{\bar k-1}|\Delta\ell_1 \rangle =O(1)$ we get
\begin{equation}
qL+O(1)\geq q\big|\langle \Delta\ell_{\bar k-1} |\Delta\ell_1 \rangle\big|
=\big( |\Delta\ell_1 |+O(1)\big)\langle q^{\bar k-1}|\Delta\ell_1 \rangle\ .
\end{equation}
Since $q>1$ Jensen's inequality yields
\begin{equation}
q^{\langle\bar k|\Delta\ell_1 \rangle}\leq\langle q^{\bar k}|\Delta\ell_1 
\rangle\leq q\frac{qL+O(1)}{|\Delta\ell_1 |+O(1)}\ .
\end{equation}
This leads to
\begin{equation}\label{erg1}
\langle\bar k|\Delta\ell_1 \rangle\leq 2+\frac{\ln\big( L/|\Delta\ell_1 |
+o(1)\big)}{\ln q}\ .
\end{equation}
Again with Jensen's inequality we have
\begin{equation}
\big\langle\ln L/|\Delta\ell_1 |\big\rangle\leq\ln\big( L\,\langle 1/|
\Delta\ell_1 |\rangle\big)
\end{equation}
Since with Lemma \ref{lemma3}, $\big\langle 1/|\Delta\ell_1
|\big\rangle =O(L^{-1/2})$ we get
\begin{equation}\label{erg2}
\langle\bar k\rangle \leq O(1)+\frac{\ln L}{2\ln q}\big( 1+o(1)\big)\ .
\end{equation}
Therefore, taking the expected value w.r.t.\ $\Delta\ell_1$, the
expected total time spent in the amplification loop is of order $L\ln
L$. Together with Lemma \ref{lemma1} this finishes the proof.\hfill
$\Box$

  \subsection{Proof of Lemmas}

\textbf{Proof of Lemma \ref{lemma2}.} 
In the following we analyze the distributions of the random
variables $t_4$ and $t_5$ to get the time evolution of
$\Delta\ell_k$. Let $\tau_n$ be the (random) time it takes for a jam
of $n$ particles to dissolve. With this
\begin{equation}\label{t4t5}
t_4 =L/2+\tau_{N_B} +E_B^k\ ,\qquad t_5 =\big( L -\Delta\ell_k \big) /2
+\tau_{N_A} +E_A^k\ .
\end{equation}
Here $N_A$ ($N_B$) denotes the number of
$A$ ($B$) particles that entered the system up to time $t_3$ ($t_2$)
before blocking, including $\Delta\ell_k$. $E_A$ ($E_B$) denotes
the number of time steps where the $A$ ($B$) jam is dissolved,
i.e. site $L$ ($1$) is empty before $t_4$ ($t_5$), when the respective marked 
particle exits. These are fluctuations and may lead to single $B$ ($A$) 
particles that enter the system due to lack of blockage. We call such 
particles discrepancies and below we show that their expected number is 
bounded independent of the system size $L$. Apart from that, the boundary site 
in a jam is always occupied and
particles are annihilated with probability $\beta$. So the time
$\tau \in\{1,2,\ldots \}$ for one particle to leave the jam is a $Geo 
(\beta )$
geometric random variable with
\begin{eqnarray}\label{geometric}
P\big(\tau =k\big) =\beta (1-\beta )^{k-1}\ ,\qquad \langle\tau\rangle
=1/\beta\ .
\end{eqnarray}
Let $\tau^i$ for $i=1,\ldots ,n$ be $n$ independent copies of $\tau$ and
 $n\in\mathbb{N}$ an independent random variable. Then for
\begin{eqnarray}
\tau_n =\sum_{i=1}^n \tau^i \quad\mbox{we have}\quad\langle\tau_n
\rangle =\langle n\rangle\langle\tau\rangle =\langle n\rangle /\beta\ .
\end{eqnarray}
In (\ref{t4t5}) $n=N_A$ or $N_B$ is a $Bi(t,\alpha )$ binomial random variable 
with
\begin{eqnarray}\label{bernoulli}
P\big( n =k\big) ={t\choose k}\alpha^k (1-\alpha )^{t-k}\ ,\quad k\in\{ 0,
\ldots ,t\}\ ,\qquad \langle n\rangle =\alpha t\ .
\end{eqnarray}
For $n=N_A$ it is $t =\Delta\ell_k /2+t_3 -B_A^k$ and for $n=N_B$, $t
=t_2 -B_B^k$, where $B_A$ ($B_B$) is the number of time steps where
the entrance of $A$ ($B$) is blocked by singular $B$ ($A$) discrepancies.
With (\ref{bernoulli}) we have
\begin{eqnarray}\label{nanb}
\langle N_A \rangle &=&\big(\Delta\ell_k /2+t_3 -\langle B_A^k \rangle\big)\,
\alpha =\big(L+\Delta\ell_k \big)\,\alpha /2-\alpha\langle B_A^k \rangle\ ,
\nonumber\\
\langle N_B \rangle &=& \big(t_2 -\langle B_B^k \rangle\big)\,\alpha 
=\big(L-\Delta\ell_k \big)\,\alpha /2-\alpha\langle B_B^k \rangle\ ,
\end{eqnarray}
where all expected values are conditioned on $\Delta\ell_k$. According to 
(\ref{geometric}), dividing (\ref{nanb}) by
$\beta$ yields $\langle\tau_{N_A}\rangle$ and $\langle\tau_{N_B}\rangle$.
Using this and (\ref{nanb})
the average value of $\Delta\ell_{k+1} =2(t_5 -t_4 )$ conditioned on
$\Delta\ell_k$ is given by
\begin{equation}\label{drift}
\big\langle \Delta\ell_{k+1} \,\big|\,\Delta\ell_k
\big\rangle = \Big( 2\frac\alpha\beta -1\Big)\,\Delta\ell_k +2\langle E_A^k 
-E_B^k \rangle -2\frac\alpha\beta \langle B_A^k -B_B^k \rangle\ .
\end{equation}
This is equal to (\ref{cycle}) with
\begin{equation}\label{ck}
C_k =2\langle E_A^k {-}E_B^k \rangle -2\frac\alpha\beta \langle B_A^k {-}B_B^k 
\rangle =2\langle E_A^k {-}E_B^k \rangle -2\frac{\alpha^2}{\beta^2} \langle 
E_A^{k-1} {-}E_B^{k-1} \rangle
\end{equation}
since the blocking of $A$ particles is caused by discrepancies of the previous 
cycle, i.e. $\langle B_A^k \rangle =\frac\alpha\beta \langle 
E_A^{k-1}\rangle$ and analogous for $B$ particles. (\ref{amp}) follows 
directly by induction.\\
In order to finish the proof it suffices to show that $\langle E_A^k \rangle$ 
and $\langle E_B^k \rangle$ are bounded independent of $L$ for all $k$. 
Recall that an $A$-jam is defined as a region of $A$ particles at the right 
boundary with densities $1$ on even and $1-\beta$ on odd sides. Denote by 
$M_t$, $t\geq t_2$, the number of particles in the $A$-jam at time 
$t$. If $M_t >0$ it decreases by $1$ with probability $\beta$ in each time 
step. Until the marked $A$ particle reaches the jam, say at time $t^* (L)$, 
$M_t$ increases at least by $1$ with probability $\alpha >\beta$ in each time 
step. Due to the sublattice parallel update, also an increase by two 
particles is possible. Both statements are true only modulo finite 
corrections due to discrepancies. Nevertheless, $M_t$ performs a biased 
random walk and is increasing on average. Thus it visits $0$ only finitely 
often for $t\to\infty$, and thus also for $t\leq t^* (L)$. Now
\begin{equation}
E_A^k =\big|\{ t\in [t_2 ,t^* (L)]\, :\, M_t =0\}\big|\ ,
\end{equation}
and thus $\langle E_A^k \rangle$ is bounded independent of $L$. The same 
argument holds for $\langle E_B^k \rangle$, finishing the proof.\hfill 
$\Box$\\
\\
\textbf{Proof of Lemma \ref{lemma3}.} With (\ref{t4t5}) and
(\ref{nanb}) we have
\begin{equation}
\langle t_5 \rangle =\big( L -\langle\Delta\ell_k \rangle\big) /2
+\big(L+\langle\Delta\ell_k \rangle\big)\,\alpha /(2\beta )-\alpha\langle 
B_A^k \rangle /\beta +\langle E_A^k \rangle\ ,
\end{equation}
where all the expected values are conditioned on $\Delta\ell_1$. This leads 
directly to (\ref{looplength}).\\
In the initial cycle starting with the empty lattice, $t_4$ and $t_5$ are 
i.i.d.r.v's and thus with $\Delta\ell_1 =2(t_5 -t_4 )$ we get
\begin{equation}
Var (\Delta\ell_1 )=4\, Var(t_5 -t_4 )=8\, Var (t_4 )\ .
\end{equation}
Analogous to (\ref{nanb}) this is easily computed as
\begin{equation}
Var (t_4 )=\langle N_A \rangle\, Var(\tau^i )+\langle\tau^i \rangle^2 \, 
Var(N_A )\ ,
\end{equation}
where $N_A$ and $\tau_i$ are defined as in the proof of Lemma
\ref{lemma1}. Thus with $\langle N_A \rangle =\alpha L/2$, $Var (N_A
)=\alpha (1-\alpha )L/2$ and $\langle\tau^i \rangle =1/\beta$, $Var
(\tau^i )=(1-\beta )/\beta$ we get
\begin{equation}
Var (\Delta\ell_1 )=4L\,\frac{\alpha (2-\alpha -\beta )}{\beta^2}
\end{equation}
\hfill $\Box$\\
\\
\textbf{Proof of Lemma \ref{lemma1}.} Let $\Delta\ell_k >L$. Then the
low density region of $A$ particles extends over the whole lattice,
and there is only a finite number of $B$ particles (discrepancies) in
the system. Therefore an $A$-jam will form at the right boundary
blocking the entrance of $B$ particles, whereas $A$ particles will not
be blocked. The number of particles in the $A$-jam performs a biased
random walk increasing on average as explained in the proof of Lemma
\ref{lemma2}, and thus the jam will reach the left boundary in an
expected time of order $L$. At this point the density of $A$
particles on all even sides is $1$ and if there are any singular $B$ 
particles left, they will leave the system in a time of order $L$ and the 
system reaches a symmetry broken configuration as defined in
Section~2.2.\hfill $\Box$

  \subsection{Proof of Theorem 2}

Assume the system to have symmetry broken distribution $\nu_A
=\mu\otimes\delta^1_\fnull$ as defined in Section~2.2 with particle
species $B$ expelled from the system. The density profile of $A$
particles is given by the stationary measure $\mu$ for the single
species system which is known exactly \cite{Schu93}. For $L\to\infty$
this means $\rho_A(i)=1+o(1)$ for even
sites and $\rho_A(i)=1-\beta +o(1)$ for odd sites, up to boundary
effects at the left boundary with $i=O(1)$. Species $B$ is expelled
from the system and
injection of $B$ particles is only possible if site $L$ is empty.
Exact expressions given in \cite{Schu93} (equation (18)) yield
\begin{equation}\label{t2result}
\rho_A (L)=1-\Big( 1-\frac{\beta\, (1-\alpha)}{\alpha\,
  (1-\beta)}\Big)\, \Big(\frac\beta\alpha\Big)^L \ .
\end{equation}
Thus the probability that site $L$ is empty is exponentially small
in the system size, and the expected time $\langle T_2
\rangle_{\nu_A}$ until the minority species can penetrate the system
started in the broken state is
exponentially large in $L$. This is not surprising even without
knowledge of the exact expressions, since for injection of the first
$B$ particle the complete jam of $A$ particles has to be dissolved
against the drive $\alpha >\beta$ and this jam consists of the order of
$L$ particles.

\section{Simulation results}

  \subsection{Results for $T_1$}
\begin{figure}
\begin{center}
  \includegraphics[width=0.75\textwidth]{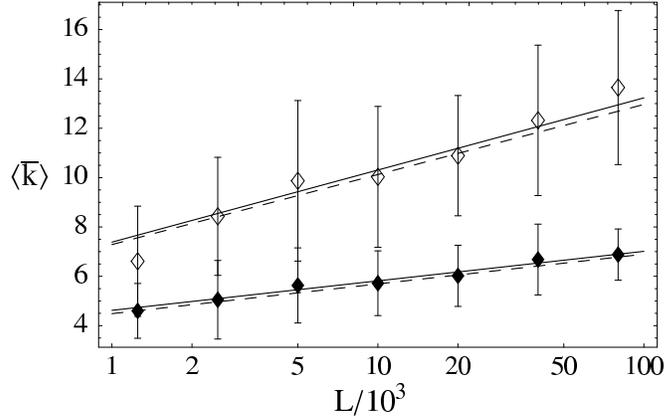}
  \caption{Simulation results for $\langle\bar k\rangle$ for different system 
sizes $L$. The data are obtained as averages over 100 realizations, the 
average values are shown as $\lozenge$ for $\alpha =0.5$, $\beta =0.4$ and 
$\blacklozenge$ for $\alpha =0.9$, $\beta =0.5$. Errors are of the
size of the symbols and the bars denote the standard deviations of the
distribution of $\bar k$. The full lines (---) give a linear fit to
the data points which agree very well with the estimate $k^* (L)$
given in (\ref{kstar}), shown as dashed lines (- - -).}
\label{fig:kbar}
\end{center}
\end{figure}

In the proof of Theorem \ref{theo1} we identified a bound on the
expected number of cycles $\langle\bar k\rangle$ until the
amplification loops stops, which is called $k^*$ in the
following. This bound grows like $k^* (L)\sim C+\ln L/(2\ln q)$ with
increasing $L$ (\ref{erg2}). To estimate 
the constant $C$ we replace $\big\langle 1/|\Delta\ell_1 |\big\rangle$ in 
(\ref{erg1}) by $1/\sqrt{Var(\Delta\ell_1)}=4L\,\frac{\alpha (2-\alpha 
-\beta)}{\beta^2}$ as given in (\ref{l1}) and get
\begin{equation}\label{kstar}
k^* (L)=\frac{\ln L}{2\ln q} +2+\frac{\ln (\beta^2/(2\alpha (2{-}\alpha 
{-}\beta ))}{2\ln q}\ .
\end{equation}
With this choice of $C$, $k^* (L)$ is no longer a strict upper bound
but in very good agreement with simulation results for $\langle\bar
k\rangle$, as can be seen in Figure \ref{fig:kbar}. In particular the
simulation data show the same logarithmic growth with prefactor $(2\ln
q)^{-1}$ as $k^* (L)$, so in this sense the rigorous bound of Theorem
\ref{theo1} is sharp.

As can be seen in Figure \ref{fig:stages} (bottom) the number of
cycles needed in each realization depends heavily on the initial
fluctuation $\Delta\ell_1$, which
is a random quantity even in the limit $L\to\infty$. Thus we do not expect a 
law of large numbers for $\bar k$, which is also supported by Figure 
\ref{fig:kbar}. The bars indicate the standard deviation of the distribution 
of $\bar k$ which is more or less independent of the system size.

  \subsection{Stationary results}

\begin{figure}[t]
\begin{center}
\raisebox{10mm}[0mm]{\includegraphics[width=0.45\textwidth]{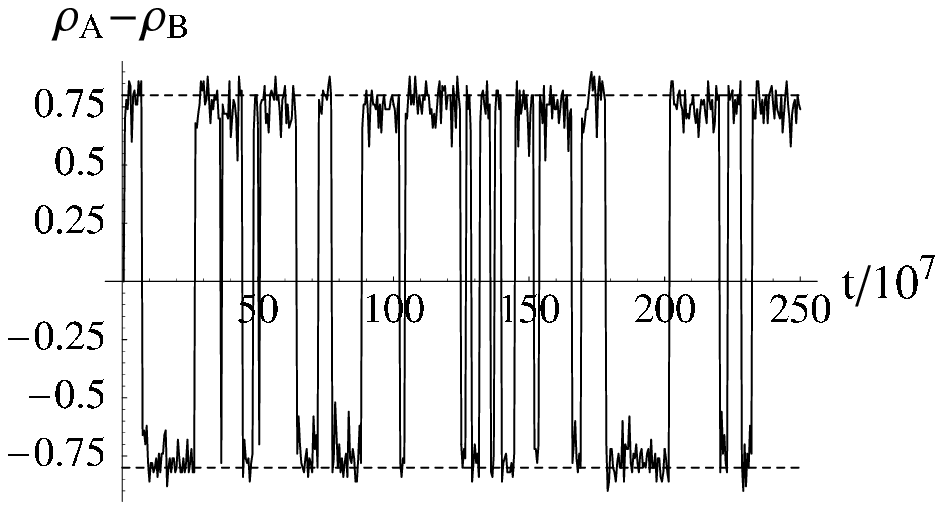}}\qquad
\includegraphics[width=0.45\textwidth]{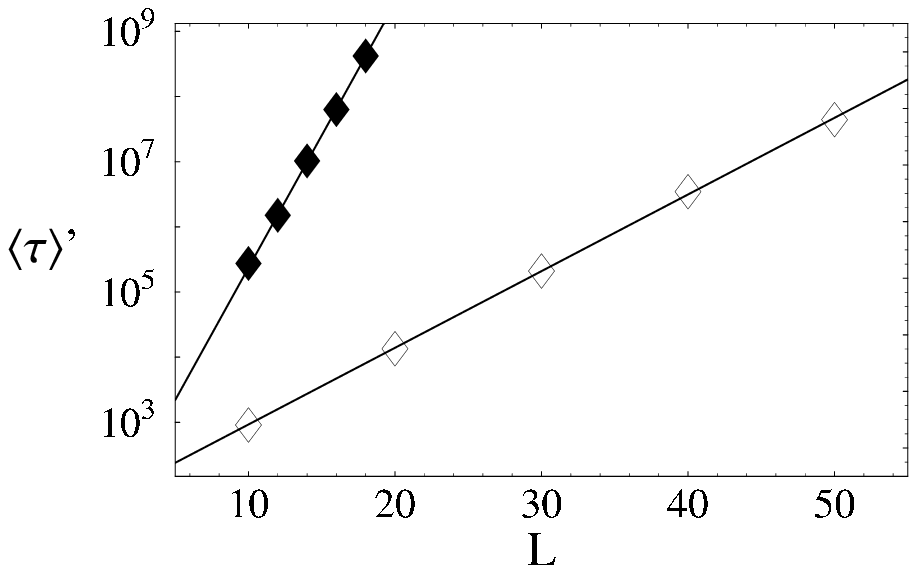}\\[5mm]
  \includegraphics[width=0.48\textwidth]{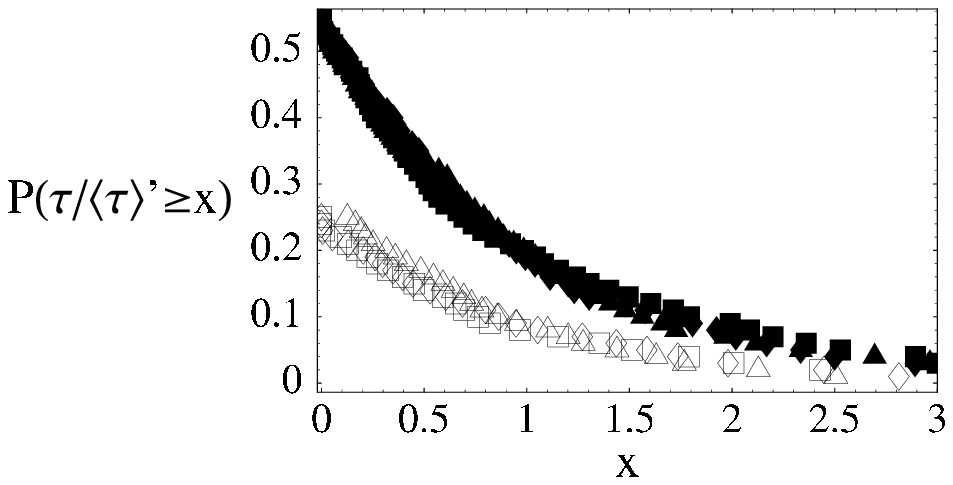}\hfill  
\includegraphics[width=0.48\textwidth]{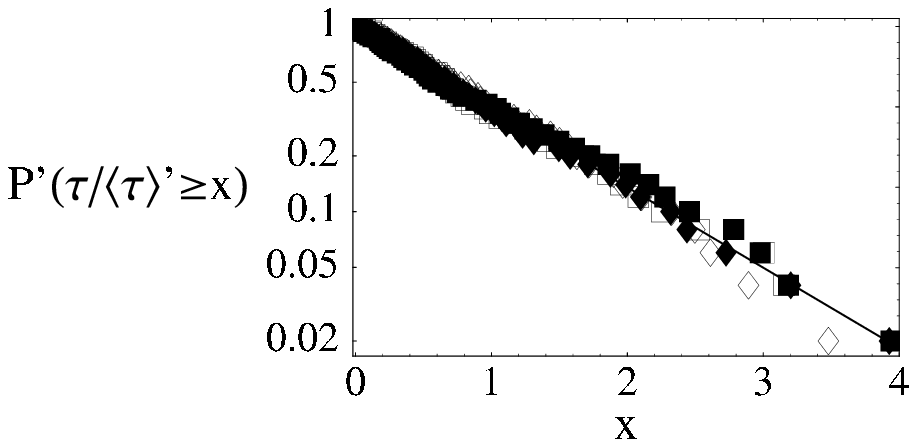}
\caption{Study of the flip time distribution. The upper left plot shows 
$\rho_A -\rho_B$ as a function of time for $\alpha =0.5$ and $\beta =0.4$ and 
$L=50$. The other plots show data for $\alpha =0.5$, $\beta =0.4$ (unfilled 
symbols) and $\alpha =0.9$, $\beta =0.5$ (black symbols). The average values 
of the relevant flip times $\langle\tau\rangle '$ (see text) are shown to be 
exponentially increasing in $L$ in the upper right plot. The full
lines give a linear fit to the data points. The lower left shows 
the cumulative tail of the distribution of normalized flip times 
$\tau /\langle\tau\rangle '$ for $L=30$ ($\bigtriangleup$), $L=40$ 
($\lozenge$), $L=50$ ($\square$) and $L=14$ ($\blacktriangle$), $L=16$ 
($\blacklozenge$), $L=18$ ($\blacksquare$). The same symbols apply in the 
lower right, a logarithmic plot of the cumulative tail of the renormalized 
distribution of relevant flip times. The full line denotes an exponential 
distribution with parameter $1$.}
\label{fig:brokenprof}
\end{center}
\end{figure}

The stationary dynamics consists mainly of flipping between the two
symmetry broken states which are close to $\nu_A$ and $\nu_B$. We
define the flip times $\tau$ to be the times between two
consecutive sign changes of $\rho_A -\rho_B$, where $\rho_A
=\frac1L\sum_{i=1}^L \eta_A (i)$ is the (time dependent) number of $A$
particles in the system normalized by $L$, and $\rho_B$ is defined
analogously. A typical trajectory of $(\rho_A -\rho_B )(t)$ is shown
in Figure \ref{fig:brokenprof} (upper left), showing that
there is a clear timescale of flipping between the two states. The lower left 
plot of Figure \ref{fig:brokenprof} shows simulation data for the cumulative 
tail of the distribution of the random variable $\tau$. This distribution 
clearly consists of two parts, one of which are small flip times $\tau =O(1)$ 
which result from fluctuations during a single transition between the two 
symmetry broken states. The relevant flip times are the ones that increase 
with the system size marking the life time of the symmetry broken states. In 
the upper right plot of Figure \ref{fig:brokenprof} the average value of 
these relevant flip times, denoted by $\langle\tau\rangle '$ is shown to 
increase exponentially in the system size as $\langle\tau\rangle '\sim
z^L$. For $\alpha =0.5$, $\beta =0.4$ we measure $z=1.31\pm 0.02$ and
for $\alpha =0.9$, $\beta =0.5$ 
$z=2.55\pm 0.08$. Both values are larger than $\alpha /\beta$, respectively, 
consistent with the result (\ref{t2result}) for $T_2$, which is a lower bound 
for the relevant flip times. Normalizing the data by $\langle\tau\rangle '$ 
in the lower two plots of Figure \ref{fig:brokenprof} results in a data 
collapse for the extensive part of the distribution. The non-extensive part 
collapses to a jump at $x=0$ in the cumulative tail (lower left plot), 
showing that a substantial fraction of sign changes have flip times of 
$O(1)$. The lower right shows a logarithmic plot of the tail of the 
renormalized extensive part of the distribution, i.e.\ the distribution of 
relevant flip times (denoted by $P'$) showing good agreement with an 
exponential distribution of parameter $1$. Therefore we conclude that the 
relevant flip times have an exponential distribution, with average value 
increasing exponentially in $L$.

\section{Dynamics on the transition line}

\begin{figure}
\begin{center}
  \includegraphics[width=0.48\textwidth]{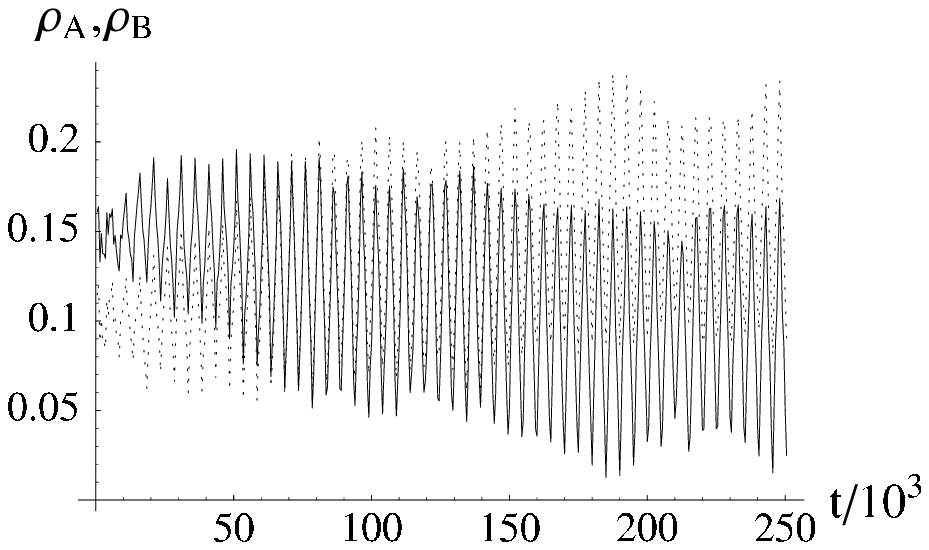}\hfill  
\includegraphics[width=0.48\textwidth]{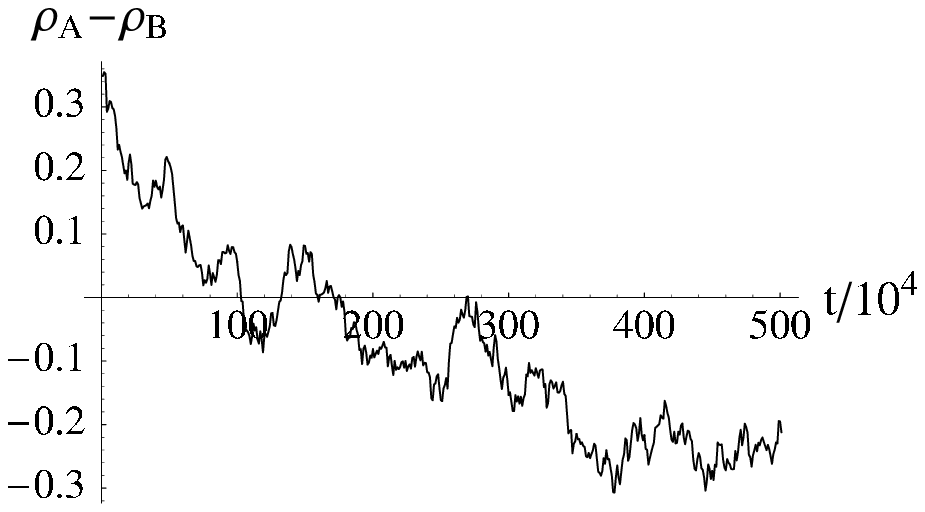}\\[5mm]
  \includegraphics[width=0.48\textwidth]{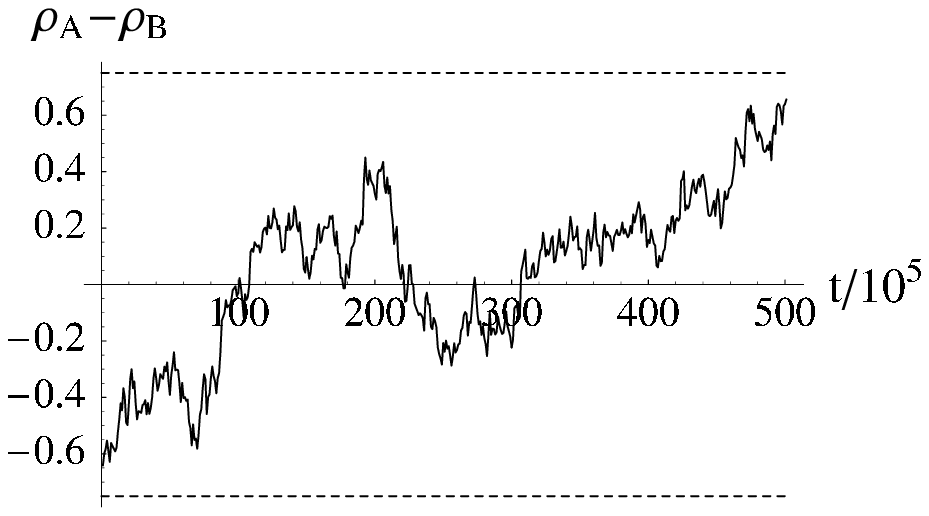}\hfill  
\includegraphics[width=0.48\textwidth]{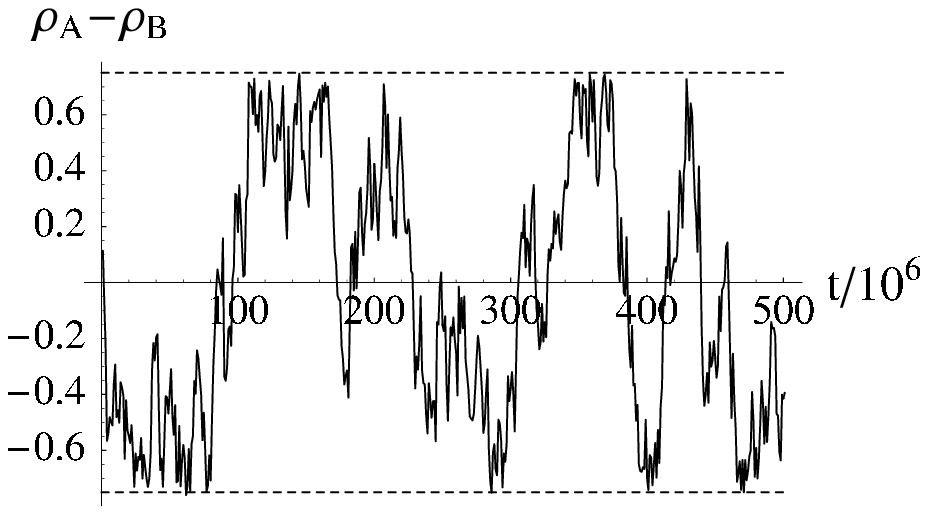}\\[5mm]
\caption{Density profiles for $\alpha =\beta =0.5$ on different time scales. 
The upper left plot shows $\rho_A$ (full line ---) and $\rho_B$
(dashed line - - -). The other plots show the difference $\rho_A
-\rho_B$. The plots are stationary samples and do not start at $t=0$,
the axes only indicate the time scale.}
\label{fig:tlineprof}
\end{center}
\end{figure}

For the borderline case $\alpha =\beta$ the dynamics of the system can still 
be effectively described in terms of boundary jams and low density regions. 
The cyclic behavior can be observed (see Figure \ref{fig:tlineprof},
upper left), but fluctuations are larger since the end of a jam is
diffusing and the cycle 
lengths, though still of order $L$, are strongly fluctuating. But according 
to (\ref{drift}) there is no amplification of fluctuations during a cycle. 
Instead, $\Delta\ell_k$ is not driven towards $\pm L$ but is diffusing,
so a symmetry broken state can still be reached within $O(L^2)$
cycles, and thus $T_1 =O(L^3)$. On the other hand, when the system
is in one of the symmetry broken states, the length of the jam of
the majority species is only diffusing. So it dissolves in a time of
only $T_2 =O(L^2)$, which is the lifetime of a symmetry broken state
for $\alpha =\beta$. Thus, no symmetry breaking takes place in this
case. Instead, for large $L$ a typical configuration is taken from a
cycle, consisting of jams with diffusing length and of low density
regions for both species. An average over many realizations leads to
an approximately linear stationary density profile as shown in Figure
\ref{fig:phasediagram} (lower left). Further, for $\alpha =\beta$
site $2$ ($L$) is occupied by $A$ particles for approximately half
of a cycle length with probability $\alpha$ ($1$), leading to
the stationary densities $\rho_A (2)=\alpha /2$ and $\rho_A (L)
=1/2$. For odd sides an analogous argument yields $\rho_A (1)=0$ and
$\rho_A (L{-}1) =(1{-}\alpha )/2$, which agrees well with Figure
\ref{fig:phasediagram}.

\begin{figure}
\begin{center}
\includegraphics[width=0.75\textwidth]{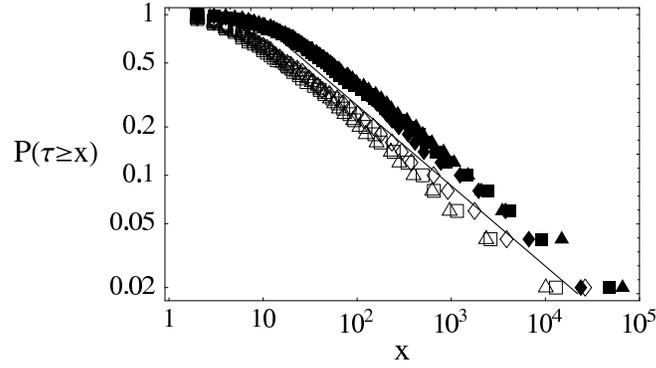}
  \caption{Double logarithmic plot of the cumulative tail of the distribution 
of flip times $\tau$. Data for $\alpha =\beta =0.9$ are shown as 
$\blacktriangle$ ($L=3200$), $\blacklozenge$ ($L=6400$), $\blacksquare$ 
($L=12800$), corresponding unfilled symbols for $\alpha =\beta =0.5$. The 
full line indicates the inclination of a power law exponent $-0.5$.}
\label{fig:tlineflip}
\end{center}
\end{figure}

Moreover, the formation and dissolution of boundary jams for the two species 
shows interesting temporal correlations \cite{Popk04b}. In the following we 
study the distribution of flip times $\tau$ between sign changes of $(\rho_A 
-\rho_B )(t)$, analogous to the symmetry broken case. In Figure 
\ref{fig:tlineprof} we show three plots of $\rho_A -\rho_B$ against time on 
different time scales for $\alpha =\beta =0.5$. For these parameters the 
maximal difference, given by the average density in the symmetry broken 
phase, is given by $1-\beta /2=0.75$. In a time of order $L^2$ the path 
explores the whole interval $[-0.75,0.75]$ (see lower right plot), but for 
smaller time scales the paths show a self similar structure (see lower left 
and upper right plot). Therefore one observes flip times on all length scales 
and in the limit $L\to\infty$ we expect a scale free distribution of $\tau$. 
This is confirmed in Figure \ref{fig:tlineflip}, showing a double logarithmic 
plot of the cumulative tail $P(\tau\geq x)$. Data for different system sizes 
collapse without scaling and show agreement with a power law tail with 
exponent $-0.5$. For large $x$ the data deviate from this behavior, due to 
boundary effects for smaller values of $L$ as discussed above, and due to 
numerical inacurracies, using quantiles to determine the cumulative tail.

The exponent of the power law can be predicted by comparison with a random 
walk. $L(\rho_A -\rho_B)$, the difference in the number of particles
as a function of
time performs a symmetric random walk for $\alpha =\beta$ on the interval 
$[-L ,L]$. Thus we use the scaling ansatz $P(\tau >x)\sim L^\gamma 
g(x/L^2 )$. The argument $x/L^2$ follows from the scaling of first passage 
times of symmetric random walks in one dimension and the fact that
$\tau$ is the return time to the origin. The power $\gamma$ can be
fixed by the requirement that 
$\langle\tau\rangle =O(L)$, since the average return time to $0$ is inversely 
proportional to the stationary distribution at $0$. This scales as $1/L$ 
because the stationary distribution of the random walker is a linear function 
on $[-L ,L]$. We get
\begin{equation}
\langle\tau\rangle\sim\int_0^\infty L^{\gamma +2} g(x/L^2 )\frac{dx}{L^2} 
=O(L^{\gamma +2})\ ,
\end{equation}
and thus $\gamma =-1$ and a consistent ansatz is
\begin{equation}
P(\tau >x)\sim L^{-1} g(x/L^2 )\ .
\end{equation}
To cancel the $L$-dependence for $L\to\infty$ we have $g(y)\sim 1/\sqrt{y}$ as 
$y\to 0$ for the scaling function and thus for $x$ large enough
\begin{equation}
P(\tau >x)\sim 1/\sqrt{x}\quad\mbox{for }x<<L^2 \ ,
\end{equation}
giving the observed exponent $-0.5$.

\section{Conclusion}

In the present article we investigated spontaneous symmetry breaking
(SSB) in a two-species driven cellular automaton model with
deterministic bulk behavior and stochastic open boundary
conditions. We analyzed in detail the dynamical mechanisms leading to
SSB. Its main feature is a cyclic amplification of initial
fluctuations taking a time of order $L\ln{L}$, while a
traffic jam phenomenon keeps the system in a SSB state for an exponentially
large time. This lead to a proof of SSB in the thermodynamic 
limit using a simple martingale argument without further assumptions
on the rates, and to rigorous
asymptotic estimates for the relevant time scales in the broken phase.
The above mechanism is very different from the freezing-by-cooling
scenario for broken ergodicity in one-component systems \cite{Rako03}
that results from a localization of shocks \cite{Parm03,Popk03b,Evans03}.

Some dynamical and stationary properties at the phase transition
line have been predicted analytically (but not rigorously) in terms of 
boundary jams and low density regions using the picture developed for the 
discussion of the broken phase. In particular, we found an asymptotically 
scale-free distribution of flip times between sign changes in the difference
of particle numbers. The decay exponent of the distribution has been
predicted using random walk arguments and confirmed by numerical
simulations. The exact phase transition line can be predicted
correctly by a mean-field approximation. The density profiles
predicted in this way, however, differ from the numerically computed
density profiles \cite{Will05}. Therefore mean field theory is
unreliable at the phase transition line, and was not presented in this
paper, whereas the correct density profiles could be predicted in
Section 5 by a rather simple argument.

The amplification mechanism outlined above does not apply in the 
symmetric phase $\alpha <\beta$ since the formation of boundary jams,
a key ingredient for the amplification, does not work. The length of a
boundary jam is driven towards small values so the boundary sites are
not blocked and particles are injected all the time.
It is an intriguing open question whether similar mechanisms
are at work in the original bridge model of \cite{Evans95a,Evans95}. 

\subsection*{Acknowledgments}
We thank M.R. Evans and V. Popkov for useful discussions. S.G. and
G.M.S. are grateful for the hospitality of the Isaac Newton Institute
for Mathematical Sciences, where part of this work has been carried
out.

\end{document}